\documentclass[twocolumn,showpacs,preprintnumbers,amsmath,amssymb]{revtex4-1}
\usepackage{graphicx}
\usepackage{dcolumn}
\usepackage{bm}
\usepackage{color}
\usepackage{SIunits}
\begin{document}
\title{Exciton-phonon interaction in the strong coupling regime in hexagonal boron nitride}
\author{T. Q. P. Vuong,$^{1}$ G. Cassabois,$^{1,\ast}$ P. Valvin,$^{1}$ S. Liu,$^{2}$ J. H. Edgar,$^{2}$ B. Gil$^{1}$}
\affiliation{$^{1}$Laboratoire Charles Coulomb, UMR 5221 CNRS-Universit\'e de Montpellier, 34095 Montpellier, France\\$^{2}$Department of Chemical Engineering, Kansas State University, Manhattan, Kansas 66506, USA}
\date{\today}
\begin{abstract}
The temperature-dependent optical response of excitons in semiconductors is controlled by the exciton-phonon interaction. When the exciton-lattice coupling is weak, the excitonic line has a Lorentzian profile resulting from motional narrowing, with a width increasing linearly with the lattice temperature $T$. In contrast, when the exciton-lattice coupling is strong, the lineshape is Gaussian with a width increasing sublinearly with the lattice temperature, proportional to $\sqrt{T}$. While the former case is commonly reported in the literature, here the latter is reported for the first time, for hexagonal boron nitride. Thus the theoretical predictions of Toyozawa [Progr. Theor. Phys. \textbf{20}, 53 (1958)] are supported by demonstrating that the exciton-phonon interaction is in the strong coupling regime in this Van der Waals crystal.
\end{abstract}
\maketitle
The electronic and optical properties in semiconductors are tightly determined by the lattice vibrations. The influence of the electron-phonon interaction emerges at zero temperature, where the zero-point fluctuations induce a renormalization of the bandgap energy, which can be a significant fraction of the unperturbed bandgap \cite{cardona,giustino}. On raising the temperature, the energy of the bandgap usually presents a monotonous decrease \cite{odonnell}, although the opposite variation also occurs in compounds such as lead chalcogenides \cite{salt}, or copper chloride \cite{gobel}.   

The temperature dependence of the excitonic optical response in semiconductors comprises an energy shift, following the modification of the bandgap energy \cite{odonnell,salt,gobel}, accompanied by a broadening of the excitonic resonance due to exciton-phonon collisions. At room temperature, this phenomenon overpasses the excitonic binding energy in many semiconductors used for optoelectronic applications \cite{rudin}. In his pioneering paper devoted to the line-shapes of the exciton absorption bands, Toyozawa pointed out the existence of two regimes for the exciton-phonon interaction \cite{toyozawa}.

The first one is the so-called strong coupling regime encountered for an efficient electron-phonon coupling, a large effective mass or a high temperature. In this case, the excitonic linewidth reflects the statistical distribution of the scattering potentials created by the fluctuations of the phonon field during lattice vibrations. Since the amplitude of the lattice distortions scales like the square root of the temperature in the harmonic approximation, the probability distribution function acquires a width $\Delta$ proportional to $\sqrt{T}$. The Gaussian distribution of the scattering potentials eventually translates into a Gaussian excitonic line with a width $\Delta$ increasing as $\sqrt{T}$.

The weak coupling regime prevails under opposite physical conditions, i.e. a low electron-phonon coupling, a small effective mass or a low temperature. Toyozawa demonstrated that the excitonic line acquires a Lorentzian profile with a width increasing linearly with temperature. In fact, such a difference results from motional narrowing for the exciton-phonon collisions. In the weak coupling regime, the product of the fluctuation amplitude $\Delta$ and fluctuation correlation time $\tau_c$ is smaller than one in unit of $\hbar$, so that the excitonic line becomes Lorentzian with a width reduced by a factor $\Delta\tau_c/\hbar$ compared to the Gaussian limit. In other words, the excitonic linewidth is given by $\Delta^2\tau_c/\hbar$ leading to a phonon-assisted broadening increasing linearly with temperature.

Since the theoretical predictions of Toyozawa in 1958, only the weak coupling regime was reported in the literature \cite{rudin}, even in semiconductors with a large bandgap and/or a strongly bound exciton, where the large effective mass is favorable to the strong coupling regime \cite{toyozawa}, such as solid xenon \cite{reilly}, sodium salts \cite{miyata}, copper chloride \cite{kaifu}, organic compounds \cite{tomioka,Jaggregates}, oxydes \cite{frohlich,bayer}, alkali halides \cite{beerwerth}, transition metal dichalcogenides \cite{TMD} or perovskites \cite{perovskite}. In contrast, both Gaussian and Lorentzian profiles are usual features in various domains, like nuclear magnetic resonance \cite{rmn}, dephasing in atomic physics \cite{sagi}, or spectral diffusion in semiconductor quantum dots \cite{berthelot}.

We demonstrate here that hexagonal boron nitride (hBN) provides a text-book example for the strong coupling regime of the exciton-phonon interaction.  We show that the emission lines have a Gaussian profile with a linewidth increasing as $\sqrt{T}$. We highlight the universal character of this phenomenology observed in hBN samples fabricated in different growth facilities. At low temperature, the phonon-assisted broadening results from quasi-elastic acoustic phonon scattering, involving the acoustic modes associated with the rigid in-phase motion of adjacent layers. Above 50K, the line-broadening is dominated by the inelastic scattering by the layer breathing modes corresponding to the low-energy optical phonons with an out-of-phase vibration of adjacent layers.

\begin{center}
\begin{figure}[t]
\includegraphics[width=0.4\textwidth]{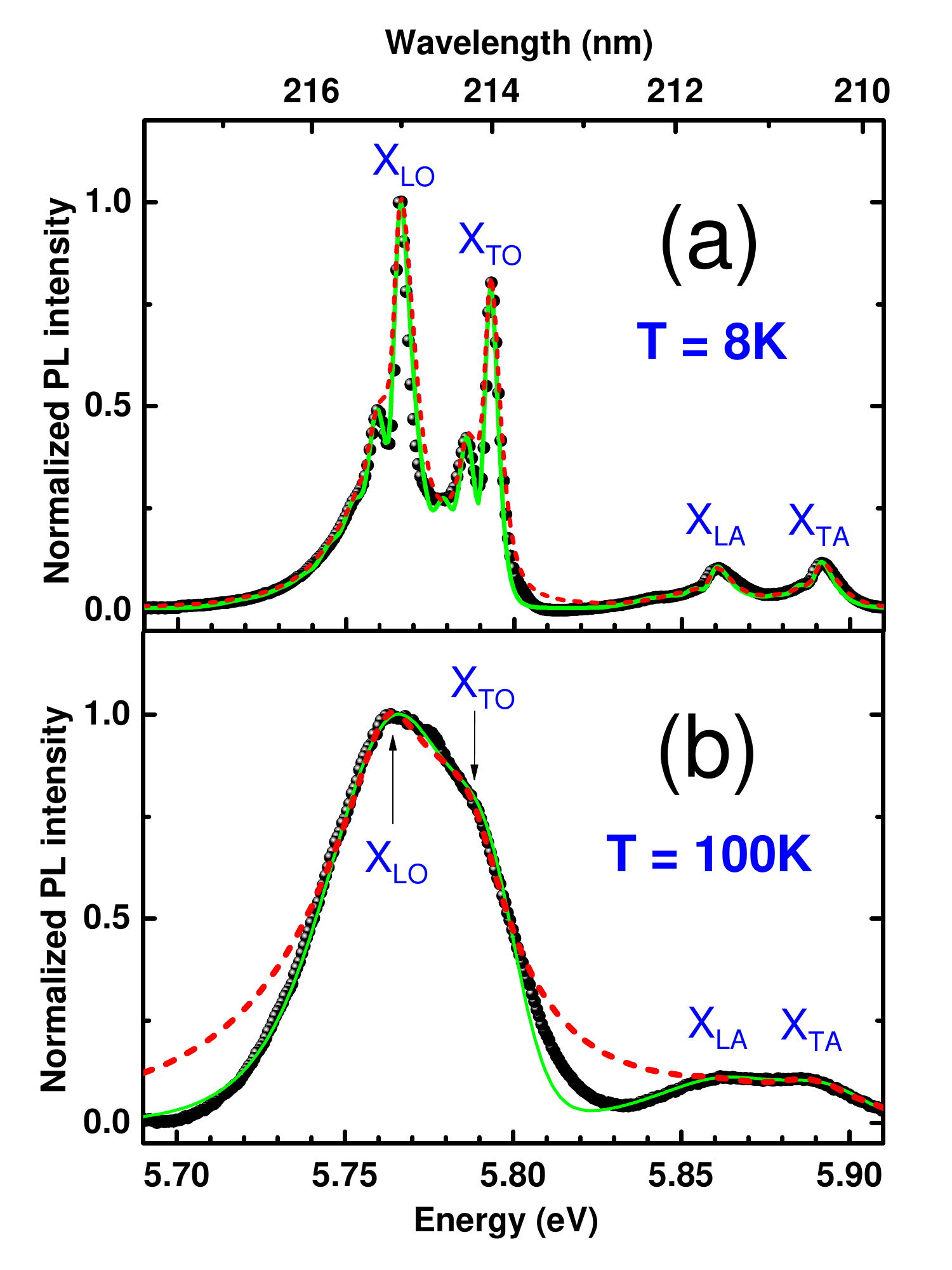}
\caption{Photoluminescence (PL) spectrum in hBN in the deep ultraviolet at 8K (a), and at 100K (b), showing the four dominant phonon replicas $X_\mu$ with $\mu$ the phonon type: experimental data (symbols), theoretical fit with a Gaussian line (green solid line), and with a Lorentzian line (dashed red line).}
\label{fig1}
\end{figure}
\end{center}

The following considerations were taken to collect a low temperature photoluminescence (PL) spectrum (Fig.\ref{fig1}(a)) from bulk hBN. In this wide bandgap semiconductor, the intrinsic optical response lies in the deep ultraviolet around 6 eV \cite{cassaboisPhot}. In order to perform above bandgap excitation at 6.3 eV, we use the fourth harmonic of a cw mode-locked Ti:Sa oscillator with a repetition rate of 82 MHz. A commercial hBN crystal from HQ Graphene is placed on the cold finger of a closed-cycle cryostat for temperature-dependent measurements. An achromatic optical system couples the emitted signal to our detection system, where the PL signal is dispersed in a f=500 mm Czerny-Turner monochromator, equipped with a 1800 grooves/mm grating blazed at 250 nm, and recorded with a back-illuminated CCD camera (Andor Newton 920), with a quantum efficiency of 50 \%, over integration times of 1 min.

The PL spectrum in bulk hBN (Fig.\ref{fig1}(a)) is composed of many emission lines, reflecting the various paths for recombination assisted by phonon emission in this indirect bandgap semiconductor. More specifically, the four peaks at 5.765, 5.79, 5.86, and 5.89 eV correspond to the recombination assisted by the emission of one LO, TO, LA and TA phonon, respectively. The identification of these phonon replicas $X_{\mu}$ (with $\mu$=LO, TO, LA, TA) was performed by means of two-photon excitation spectroscopy revealing the phonon energy ladder up to the indirect exciton energy at 5.955 eV, together with the phonon symmetries in polarization-resolved measurements \cite{cassaboisPhot,vuong2D}. On the low-energy side of each of the aforementioned lines, there are sidebands which extend over tens of meV, and which arise from overtones of interlayer shear modes in hBN \cite{vuongPRB}. These lattice vibrations are specific to layered compounds since they correspond to the shear rigid motion between adjacent layers, with a characteristic energy of about 6-7 meV in bulk hBN \cite{cuzco}. The latter energy is directly observable from the doublet structure of the LO and TO replicas around 5.765 and 5.79 eV in Fig.\ref{fig1}(a).

The green solid line in Fig.\ref{fig1}(a) is a theoretical calculation of the emission spectrum in bulk hBN in the framework of the model developed in Ref.\cite{vuongPRB}. The strength of this approach reaching a quantitative interpretation of the full PL spectrum in the deep ultraviolet is the use of a single fitting parameter $\Delta$, which is identical for all types of phonon replicas, and which phenomenologically accounts for the line-broadening. In the calculations, the only parameter which depends on the phonon replica is the phonon group velocity, displaying strong variations within the different phonon branches \cite{cuzco}. The lower the phonon group velocity, the higher the visibility of the doublet structure, as for the TO phonon replica at 5.79 eV (Fig.\ref{fig1}(a)) in contrast to the LA one at 5.86 eV where the group velocity is seven times higher, so that the overtones sideband largely overlaps with the fundamental LA phonon replica, smearing out the doublet structure.

The first piece of evidence for the strong coupling regime of the exciton-phonon interaction is the Gaussian profile of the emission lines. In Fig.\ref{fig1}, the theoretical emission spectrum is calculated after convolution with either Lorentzian (red dashed line in Fig.\ref{fig1}(a\&b)) or Gaussian (green solid line in Fig.\ref{fig1}(a\&b)) functions in order to account for the phenomenological broadening of the emission lines. The superior fit of the data by the Gaussian function supports the interpretation as strong coupling. In contrast, with a Lorentzian function, with the same full width at half maximum (FWHM) for accurately fitting the central part the emission lines, one misses the key feature of the doublet at low temperature (Fig.\ref{fig1}(a)) because of the slow decay of Lorentzian wings. Similarly, at high temperature (Fig.\ref{fig1}(b)) where the doublet structure due longitudinal and transverse phonons is hardly observable, the broad emission bands centered at 5.77 and 5.88 eV decay to rapidly for being accounted for by a Lorentzian function (red dashed line in Fig.\ref{fig1}).

A Gaussian lineshape is a necessary but insufficient condition for demonstrating the strong coupling regime predicted by Toyozawa \cite{toyozawa}: Gaussian profiles are not unusual. In fact, a Gaussian emission line is a widely encountered situation in optical spectroscopy as soon as disorder or defects induce inhomogeneous broadening. In that case, a Gaussian lineshape is the signature of extrinsic broadening processes leading to a sample-dependent emission spectrum. The situation here is drastically different: the same peak shape emission spectrum above 5.7 eV appears repeatably in bulk hBN at low temperature, as illustrated for instance by Ref.\cite{schue}. In the present case, the quantitative analysis of the temperature-dependent measurements in two samples grown in different laboratories further confirm the intrinsic origin of the Gaussian broadening of the emission lines in bulk hBN. Let us finally mention that the assumption of cumulative broadening as a function of the overtone index in Ref.\cite{vuongPRB} also points out the exciton-phonon coupling as the broadening mechanism in bulk hBN since the phonon-assisted broadening increases with the order of the electron-phonon interaction process. In summary, the identification of Gaussian emission lines in the PL spectrum of bulk hBN at low temperature comes along with strong indications as arising from the exciton-phonon interaction. Even though we have no absolute proof for a negligible residual inhomogeneous broadening in hBN at low temperature, the additional comparison of Gaussian and Lorentzian functions at 100K (Fig.\ref{fig1}(b)) definitely rules out this alternative mechanism at high temperature. A Gaussian function remains superior for fitting the emission lines when they are much broader due to phonon-assisted scattering.
\begin{center}
\begin{figure}[t]
\includegraphics[width=0.5\textwidth]{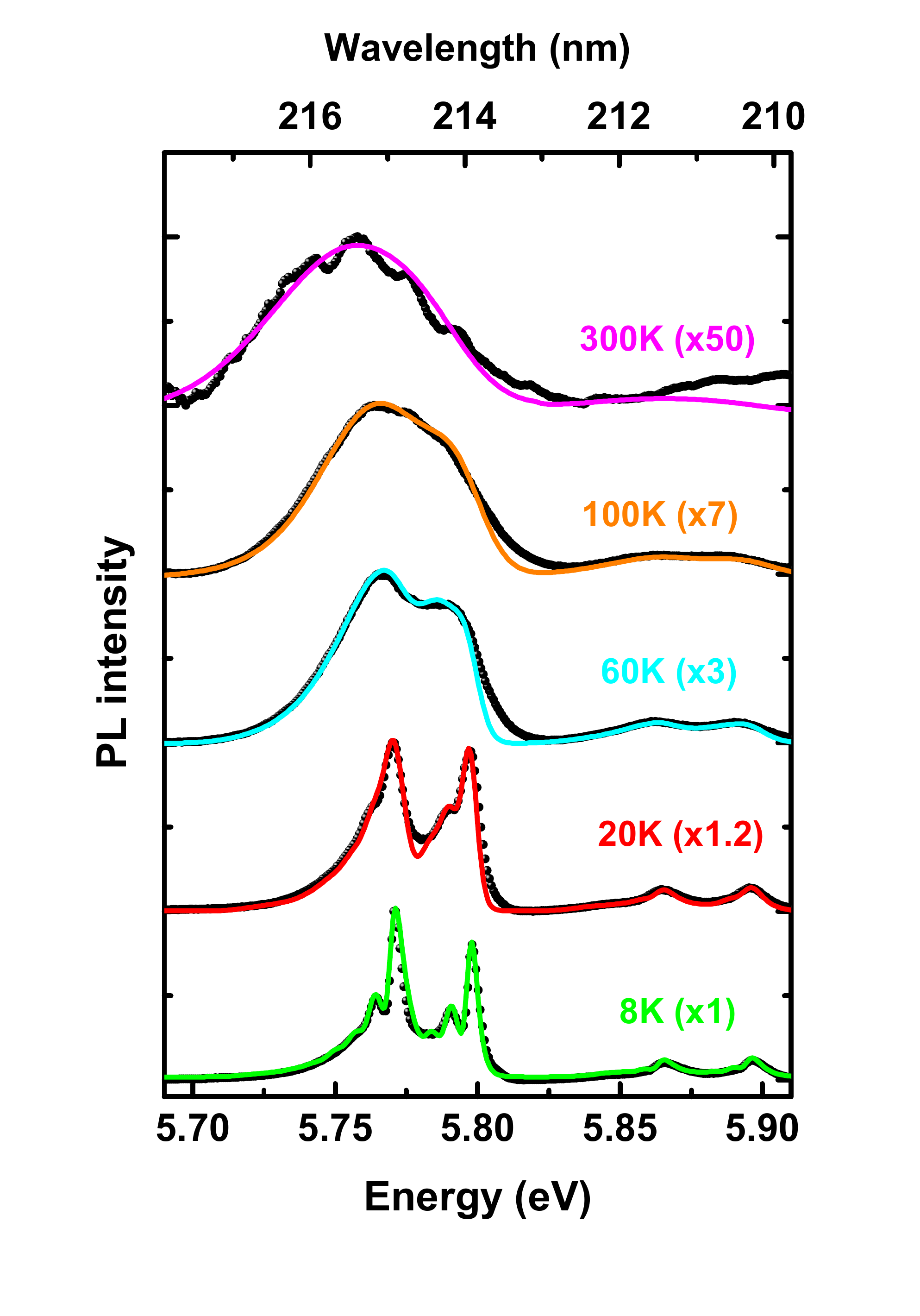}
\caption{Temperature dependence of the PL spectrum in hBN as a function of temperature, from 8 to 300K: experimental data (symbols), theoretical fit (solid line). The full width at half maximum $\Delta$ of the phonon replicas is the same for the four types of phonon replicas (as shown in Fig.\ref{fig1}), and $\Delta$ is the only varying parameter as a function of temperature.}
\label{fig2}
\end{figure}
\end{center}

The second piece of evidence for the strong coupling regime of the exciton-phonon interaction relies on PL measurements from 10K to 300K. The temperature-dependent study of the PL spectrum in hBN is reported in Fig.\ref{fig2} where a few temperatures have been selected for the sake of clarity. On increasing the temperature, we observe the progressive broadening of the various phonon replicas, leading first to a reduced visibility of the doublet fine-structure of the LO and TO phonon replicas. Above 20K, it is hardly observable, thus indicating that the FWHM of the emission lines becomes larger than the doublet splitting of the order of 6-7 meV. Above 100K, we note that the LO and TO phonon replicas separated by roughly 30 meV are no longer spectrally resolved, and that the merged emission band of these replicas is almost symmetric. From 100K up to room temperature, the modifications of the PL spectrum are less dramatic even if it becomes more and more difficult to distinguish the emission band coming from the LA and TA phonon replicas at high energy, because of the reduced signal-to-background ratio. The above phenomenology therefore suggests that the thermally-induced broadening is rapidly growing in a first stage, followed by smoother variations up to room temperature.

A quantitative analysis of the phonon-assisted broadening in bulk hBN has been implemented by extending the approach which was developed at low temperature in Ref.\cite{vuongPRB} for the interpretation of the low-energy sidebands due to the interlayer shear modes. Again, as already highlighted above, the strength of this model is the use of a single parameter $\Delta$ for reproducing the full PL spectrum of bulk hBN in the deep ultraviolet. Once the low temperature PL spectrum is accurately reproduced with the calculated values of the phonon group velocities \cite{vuongPRB}, the temperature-dependent PL spectrum is adjusted by only varying $\Delta$. This robust fitting procedure allows us to fairly account for the complete set of measurements, as shown in Fig.\ref{fig2} where the theoretical fits are plotted as solid lines.
\begin{center}
\begin{figure}[t]
\includegraphics[width=0.5\textwidth]{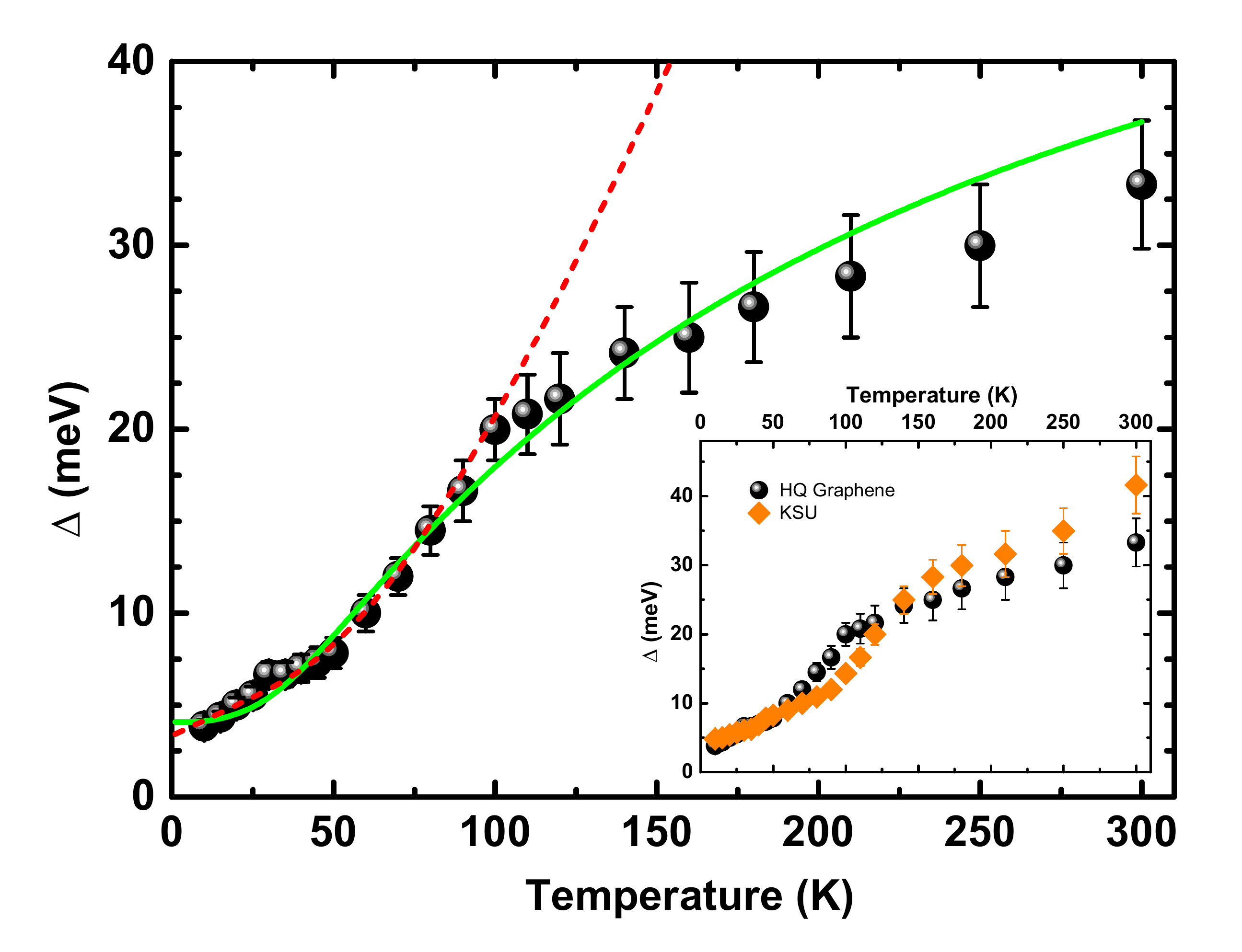}
\caption{Temperature dependence of the full width at half maximum $\Delta$ of the phonon replicas, estimated from the quantitative interpretation performed in Fig.\ref{fig2}: experimental data (symbols), theoretical fit in the strong coupling regime (green solid line), or in the weak coupling regime (dashed red line). Inset: comparison between a commercial hBN sample from HQ Graphene and a crystal grown in Kansas State University (KSU) \cite{edgar}.}
\label{fig3}
\end{figure}
\end{center}

From this analysis, we thus extract the temperature dependence of $\Delta$, corresponding to the FWHM of the phonon replicas, and which is plotted in Fig.\ref{fig3}. Above 100K, the sublinear increase of $\Delta$ as a function of temperature stems from the $\sqrt{T}$-variations of the linewidth. However, the temperature dependence is not as straightforward as a pure $\sqrt{T}$-broadening, revealing the contribution of two types of scattering processes in the phonon-assisted broadening.

The solid line in Fig.\ref{fig3} is a fit of the thermally-induced broadening according to the expression:  
\begin{equation}
\Delta=\sqrt{\Delta_A^2+\Delta_O^2}
\label{eq1}
\end{equation} 
where $\Delta_A$ and $\Delta_O$ are the broadenings due to acoustic and optical phonons, respectively. Such an expression is specific to Gaussian broadening, the convolution of two Gaussian functions of width $\Delta_A$ and $\Delta_O$ having a width $\Delta$ given by Eq.\ref{eq1}.

$\Delta_A$ arises from quasi-elastic scattering by acoustic phonons of mean energy $E_A$, and $\Delta_A$ is given by \cite{toyozawa,sumi}:
\begin{equation}
\Delta_A^2=S_AE_A\coth\left(\frac{E_A}{2k_BT}\right)
\label{eq2}
\end{equation} 
where $k_B$ is the Boltzmann constant, and $S_A$ is the strength of the coupling to phonons, which also represents the renormalization of the exciton energy due to the exciton-phonon interaction \cite{toyozawa,sumi}.

In the case of optical phonons of energy $E_O$, the inelastic nature of the exciton-phonon collisions excludes the scattering processes due to phonon emission \cite{rudin}, so that $\Delta_O$ only stems from phonon absorption, thus reading:
\begin{equation}
\Delta_O^2=S_OE_O\frac{1}{e^\frac{E_O}{k_BT}-1}
\label{eq3}
\end{equation} 

In Fig.\ref{fig3}, we reach an excellent agreement with our data by taking $E_A$=4$\pm$2 meV, $S_A$=4.5$\pm$1.5 meV, $E_O$=15$\pm$5 meV, and $S_O$=60$\pm$8 meV.

Before further discussing the specificity of phonon-assisted broadening in a Van der Waals crystal such as hBN, we aim at comparing with the standard expression used in the weak coupling regime for Lorentzian lines: the FWHM is fitted in that case by $\Gamma_0$+$aT$+$b/(\exp(E_O/k_BT)-1)$, where the last two terms account for acoustic and optical phonon broadening, respectively \cite{rudin,reilly,miyata,kaifu,tomioka,Jaggregates,frohlich,bayer,beerwerth,TMD,perovskite}. The best fit is displayed in dashed line in Fig.\ref{fig3}, and we find $\Gamma_0$=3$\pm$0.5 meV, $a$=0.1$\pm$0.02 meV/K, $b$=150$\pm$80 meV and $E_O$=25$\pm$10 meV. However, there is a large and obvious discrepancy with the data of the temperature-dependent linewidth in Fig.\ref{fig3}. Consequently, from the combined evidence of a Gaussian lineshape (Fig.\ref{fig1}) and prominent $\sqrt{T}$-dependence of the linewidth (Fig.\ref{fig3}), we thus conclude that the exciton-phonon interaction in bulk hBN is in the strong coupling regime predicted by Toyozawa. 

At low temperature, below 50K, the phonon-assisted broadening is dominated by $\Delta_A$, which results from quasi-elastic acoustic phonon scattering with phonons of mean energy $E_A$=4$\pm$2 meV. This means that the involved phonons mostly originate from the so-called ZA branch close to the zone-center \cite{cuzco}. These acoustic modes are associated with out-of-plane atomic displacements with a rigid in-phase motion of adjacent layers.

Very importantly, we stress that there is no off-set term in Eq.\ref{eq1}, equivalent to $\Gamma_0$ for Lorentzian lines \cite{rudin}. Here, the finite value of the linewidth at low temperature arises from $\Delta_A$ itself because of the quasi-elastic nature of the scattering processes involving ZA phonons. Because of the indirect nature of the bandgap, no radiative broadening contributes to the zero-temperature limit of the linewidth. In bulk hBN, the latter value is solely determined by the intrinsic process of phonon scattering without any other contribution. This explains why the same emission profiles is systematically detected at cryogenic temperatures in hBN samples fabricated in different growth facilities \cite{schue}. In order to be more quantitative on this point, we plot in the inset of Fig.\ref{fig3} the comparison of our commercial hBN crystal from HQ Graphene with a sample grown at Kansas State University (KSU) \cite{edgar}. Below 50K, $\Delta$ is exactly the same, further supporting that the zero-temperature linewidth is exclusively limited by phonon-assisted broadening in bulk hBN. The temperature-dependence of $\Delta$ in the KSU sample is fitted with $E_A$=4$\pm$2 meV, $S_A$=3.5$\pm$1.5 meV, $E_O$=15$\pm$5 meV, $S_O$=65$\pm$10 meV. These values are identical, within our fitting error, to the ones found above for the HQ Graphene sample.

We finally comment the broadening term $\Delta_O$ which dominates over 50K, with a characteristic energy $E_O$=15$\pm$5 meV. This estimation exactly matches the energy of the ZO$_1$ optical branch at the zone center, corresponding to the silent low-energy optical phonons with an out-of-phase vibration of adjacent layers \cite{cuzco}. As a matter of fact, the $\Delta_O$ phonon-assisted broadening is due to inelastic scattering by the layer breathing modes in hBN, which are specific to layered compounds. In the weak coupling regime usually encountered in all semiconductors \cite{rudin,reilly,miyata,kaifu,tomioka,Jaggregates,frohlich,bayer,beerwerth,TMD,perovskite}, the interaction with optical phonons also provides the dominant contribution to dephasing at room temperature, however the impact is much more dramatic because of the linear dependence of the broadening with temperature. Here, in contrast, exciton scattering by absorption of layer breathing modes does not induce a significant broadening of the emission lines between 100 and 300K because of the $\sqrt{T}$-variations inherent to the strong coupling regime.  

The demonstration of the strong coupling regime in bulk hBN raises the interesting question of its link with the striking bright emission \cite{watanabe04,watanabe09} despite the indirect bandgap \cite{cassaboisPhot,vuong2D}. Thanks to the efficient exciton-phonon interaction, we infer that the phonon-assisted recombination is fast enough to by-pass non-radiative relaxation, therefore leading to an intense emission in the deep ultraviolet. The strong coupling regime of the exciton-phonon interaction is indeed a fundamental effect, which appears to be connected with one of the key aspects of this material for optoelectronic applications. The strong coupling regime in hBN may originate from the large effective mass in this lamellar compound \cite{arnaud}, a large effective mass being one of the key ingredients, as discussed by Toyozawa in Ref.\cite{toyozawa}. This would mean that the specific cristallographic structure of Van der Waals crystals is beneficial for the strong coupling regime. Still, the theoretical investigation of the electron-phonon interaction in hBN is to be done, and it will be necessary for reaching a detailed understanding of the efficient exciton-phonon coupling in this material.

In summary, hBN provides a text-book example for the strong coupling regime of the exciton-phonon interaction, predicted by Toyozawa but not previously observed. The emission spectrum in the deep ultraviolet displays the two required signatures of a Gaussian emission line and $\sqrt{T}$-increase of the linewidth. We have quantitatively interpretated the temperature dependence of the linewidth on the basis of quasi-elastic scattering by acoustic phonons, and inelastic scattering by absorption of optical phonons corresponding to the layer breathing modes in hBN. The strong coupling regime and the bright emission in this indirect bandgap semiconductor are unusual properties, calling for a microscopic understanding of the exciton-phonon coupling in this Van der Waals crystal with fascinating properties.

\textbf{Acknowledgments}

We gratefully acknowledge C. L'Henoret for his technical support at the mechanics workshop, and V. Jacques for fruitful discussions. This work and the PhD funding of T. Q. P. Vuong were financially supported by the network GaNeX (ANR-11-LABX-0014). GaNeX belongs to the publicly funded \textit{Investissements d'Avenir} program managed by the French ANR agency. G.C. is member of 'Institut Universitaire de France'. The hBN crystal growth is based upon work supported by the National Science Foundation under Grant No. 1538127.

$^\ast$e-mail: guillaume.cassabois@umontpellier.fr

\end{document}